\documentclass{article}

 \usepackage[final]{neurips_2025_ml4ps}

\usepackage[utf8]{inputenc} 
\usepackage[T1]{fontenc}    
\usepackage{hyperref}       
\usepackage{url}            
\usepackage{booktabs}       
\usepackage{amsfonts}       
\usepackage{nicefrac}       
\usepackage{microtype}      
\usepackage{xcolor}         
\usepackage{wrapfig} 
\usepackage{graphicx}
\usepackage{amsmath}
\usepackage{orcidlink}

\title{Neural Deprojection of Galaxy Stellar Mass Profiles}

\author{%
    \textbf{M. J. Yantovski-Barth}$^{1,2,3}$\orcidlink{0000-0001-5200-4095} \and 
    \textbf{Hengyue Zhang}$^{4}$\orcidlink{0000-0003-4079-2447} \and 
    \textbf{Nolan Smyth}$^{1,2,3}$\orcidlink{0000-0002-8454-3015} \and
    \textbf{Connor Stone}$^{1,2,3,5}$\orcidlink{0000-0002-9086-6398}\and 
    \textbf{Martin Bureau}$^{4}$\orcidlink{0000-0003-4980-1012} \and 
    \textbf{Yashar Hezaveh}$^{1,2,3,6}$\orcidlink{0000-0002-8669-5733} \and 
    \textbf{Laurence Perreault-Levasseur}$^{1,2,3,6}$\orcidlink{0000-0003-3544-3939}
    \\
    $^{1}$Universit{\'e} de Montr{\'e}al \quad $^{2}$Ciela Institute \quad $^{3}$Mila \quad $^{4}$University of Oxford \quad \\ $^{5}$University of Toronto, Dunlap Institute \quad $^{6}$CCA, Flatiron Institute \\
    \texttt{\{michael.barth,nolan.smyth,yashar.hezaveh,}\\ \texttt{laurence.perreault.levasseur\}@umontreal.ca}\\
    \texttt{\{hengyue.zhang, martin.bureau\}@physics.ox.ac.uk}\\
}

\begin{document}

\maketitle

\begin{abstract}
  We introduce a neural approach to dynamical modeling of galaxies that replaces traditional imaging-based deprojections with a differentiable mapping. Specifically, we train a neural network to translate Nuker profile parameters into analytically deprojectable Multi Gaussian Expansion components, enabling physically realistic stellar mass models without requiring optical observations. We integrate this model into SuperMAGE, a differentiable dynamical modelling pipeline for Bayesian inference of supermassive black hole masses. Applied to ALMA data, our approach finds results consistent with state-of-the-art models while extending applicability to dust-obscured and active galaxies where optical data analysis is challenging.
\end{abstract}

\section{Introduction}\label{sec:intro}
A central open question in galaxy evolution is the origin of the tight correlation between various galaxy properties and the mass of the supermassive black hole (SMBH) at the centre of each galaxy (e.g., [1, 2, 3]). Probing this correlation requires precision mass measurements of SMBHs in a wide variety of galaxies. Dynamical modelling of cold molecular gas has produced some of the most accurate black hole mass measurements to date [4]. Since the motions of gas trace the total gravitational potential, the fundamental challenge in dynamical modelling is to disentangle the mass of stars from the mass of the SMBH. Traditionally, the following approach has been used (e.g., [5, 6]): 
\begin{enumerate}
    \item Optical observations of the target galaxy provide an estimate of the projected surface density of light from stars in the galaxy. 
    \item The optical surface density is deprojected into 3D by fitting a Multi Gaussian Expansion (MGE) model to the 2D optical surface brightness [7]:
    \begin{equation}
        \Sigma(x,y)=\sum_{i=1}^{N}\frac{\Sigma_i}{2\pi\sigma_i^2q'_{i}}\exp{\left[-\frac{1}{2\sigma_i^2}\left(x^2+\frac{y^2}{q_i'^2}\right)\right]}.
    \end{equation}
    Here, $\Sigma$ is the surface brightness and $x$ and $y$ are the major-axis and the minor-axis coordinates. The MGE fits the 2D surface brightness distribution with a sum of Gaussians, each with a different normalization $\Sigma_i$, standard deviation $\sigma_i$, and sky-projected axis ratio $q_i'$. The Gaussians are then deprojected into 3D, given an inclination $i$ and assuming axisymmetry, simply by converting $q_i'$ and $i$ into intrinsic axis ratios $q_i$. The deprojected MGE light profile is converted to a 3D mass density using a mass-to-light ratio ($M/L$), and a point mass is added to represent the black hole.
    \item The corresponding gravitational potential is converted to a velocity curve. The orbital velocity of the gas "Doppler" shifts the light it emits, which allows the velocity to be inferred from the data based on the shifted spectrum of the emitted light. Hence, the velocity curve is combined with a model for the light-emitting gas to predict the spatial variations in the spectrum of the galaxy. 
\end{enumerate}
The above model is then fitted to radio interferometer observations of the spectrum of the gas. However, this model is subject to several limitations. Optical observations of the stars must not be contaminated by dust, which absorbs the light emitted by stars (e.g., [8]). Hence, only relatively dust-free galaxies may be analysed. One of the strengths of the MGE can also become a limitation: the MGE model has a large number of parameters, which allows it to fit the light distribution to high precision. In the presence of an active galactic nucleus (AGN), whose light mimics a concentration of stars at the centre of the galaxy, the MGE model will fit the AGN and thereby underestimate the SMBH mass (e.g., [9]). Finally, in the case of an extremely bright AGN, known as a quasar, fitting any stellar light model to optical images becomes impossible, as quasars outshine their host galaxies by several orders of magnitude (e.g., [10]). In this work, we address the above limitations with a new dynamical modelling methodology which does not require optical imaging of the galaxy. 

\section{Methodology}
\subsection{Neural deprojection of the Nuker model}\label{sec:Nuker}
When optical imaging is unavailable, MGE models of stellar mass become degenerate with the SMBH mass, since the SMBH can be replaced by a central stellar overdensity. To enforce realistic stellar mass profiles, we adopt the Nuker model [11]:
\begin{equation}
\Sigma_*(r) \;=\; \Sigma_b\,2^{(\beta-\gamma)/\alpha}
\left(\frac{r}{r_b}\right)^{-\gamma}
\left[\,1+\left(\frac{r}{r_b}\right)^{\alpha}\right]^{(\gamma-\beta)/\alpha},
\label{eq: Nuker}
\end{equation}
where $\Sigma_*$ is the stellar surface mass density, and $r=\sqrt{x^2+y^2/q'^2}$ is the galactocentric radius (for a given 2D axis ratio $q'$). The free parameters of the model are the inner power law slope $\gamma$, the outer power law slope $\beta$, the transition sharpness between the inner and the outer power law $\alpha$, the break radius of the transition $r_b$, and the surface mass density at transition $\Sigma_b$. However, we cannot use the Nuker model on its own, as it lacks an analytic axisymmetric deprojection into 3D \footnote{A spherically symmetric deprojection exists [12], but since the stars at the centres of galaxies are typically axisymmetrically distributed, assuming spherical symmetry will lead to significant bias.}. To combine the realistic stellar density profile of the Nuker with the analytic deprojection of the MGE, we require a mapping between the two parametrizations. We achieve this mapping by training a neural network (NN) to predict 1D MGE components that give a mass profile that corresponds to a given set of Nuker parameters. Hence, the MGE components serve as an intermediate parametrization which facilitates the calculation of a deprojected velocity profile for the parameters of interest (Nuker). 

\subsubsection{Training Data}\label{sec: training data}
Our training set consists of $6.25e6$ simulated Nuker profiles covering typical literature parameter ranges (see Appendix \ref{Nuker_range}). We vary the parameters $\alpha,\beta,\gamma$, and $r_b$; we do not need to include $\Sigma_b$ in the training, as the true profile can be recovered simply by rescaling the Gaussian normalizations $\Sigma_i$ with $\Sigma_b$. The training set thus consists of $6.25e6$ sets of Nuker parameters $\alpha,\beta,\gamma$, $r_b$ and the corresponding mass profiles calculated from Equation \ref{eq: Nuker} with $\Sigma_b$ set to $1$.   

\subsubsection{Neural network architecture and training}\label{sec: neural net}
\begin{wrapfigure}{r}{0.58\textwidth}
\vspace{-0.1\linewidth}
    \includegraphics[width=\linewidth]{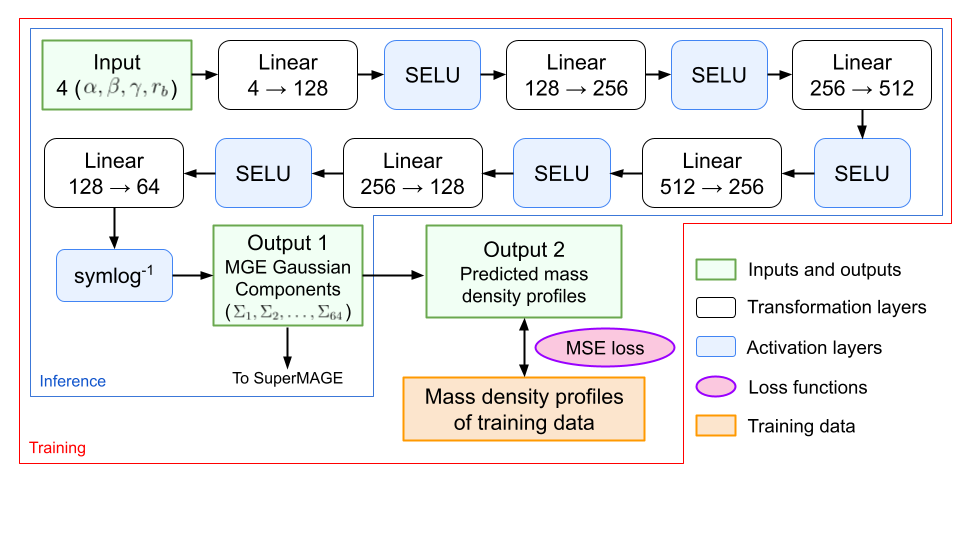}
    \vspace{-0.12\linewidth}
    \caption{A flowchart that illustrates the architecture of the NN that maps Nuker parameters to MGE parameters.}
    \label{fig: NN_arch}
    \vspace{-0.05\linewidth}
\end{wrapfigure}

Figure \ref{fig: NN_arch} represents our neural architecture. The last fully connected layer of our neural network outputs 64 values (Output 1) which represent the normalizations $\Sigma_i$ of an MGE profile with 64 Gaussian components on a fixed grid of $\sigma_i$ (see Appendix \ref{priors} for details). 
The Gaussian normalizations $\Sigma_i$ required to accurately fit most Nuker profiles span a large dynamic range, from $\sim \pm 10^{-12}$ to $\sim \pm10^3$; if we train the neural network to predict these values directly it destabilizes training. We thus apply an inverse symlog (or "symexp") activation function,
\begin{equation}
    \Sigma_i=\mathrm{symexp}(\Sigma_i')
= \mathrm{sgn}(\Sigma_i')\, 10^{-12}\!\left(10^{|\Sigma_i'|}-1\right),
\end{equation}
where $\Sigma_i'$ are the 64 values output by the final linear transformation layer and $\Sigma_i$ are the 64 MGE normalizations (Output 1 in Figure \ref{fig: NN_arch}). The symexp transformation compresses the dynamic range of $\Sigma_i'$ into $\sim \pm 0.3$ to $\sim \pm15$, which simplifies learning the weight matrix for that layer.

To train the neural network to output Gaussian normalizations (Output 1 in Figure \ref{fig: NN_arch}), we feed the 64 predicted $\Sigma$ values and the corresponding fixed grids of $\sigma_i$ and galactocentric radius $r$ into the following equation:
\begin{equation}
    \Sigma_{\mathrm{MGE}}(r)
= \sum_{i=1}^{N} \Sigma_i \,
\exp\!\left(-\frac{r^{2}}{2\sigma_i^{2}}\right),
\label{eq: radial MGE}
\end{equation}
where $\Sigma_{\mathrm{MGE}}(r)$ is the surface mass density profile of an MGE (Output 2 in Figure \ref{fig: NN_arch}). We then adopt a mean squared error (MSE) loss between the predicted $\Sigma_{\mathrm{MGE}}(r)$ and the true $\Sigma(r)$ corresponding to the Nuker profile in the training data. 

We minimize the loss function using the Adam optimizer [13] with a learning rate of $10^{-5}$. After $\approx10$ hours of training on an NVIDIA GeForce RTX 4060 TI GPU, the NN achieves $<3\%$ fractional error in predicting the mass profile. This performance is more than sufficient for our science task, given that the statistical uncertainty of the best-fitting Nuker profile is much larger than $3\%$ (see Figure~\ref{fig:nukervsmge}). Further improvements may be attained with additional training time and hyperparameter tuning. 

During inference, Equation \ref{eq: radial MGE} is bypassed, and instead the MGE parameters are fed directly into our forward modelling code (see Section \ref{sec: SuperMAGE}) which converts them to a velocity profile; the velocity profile is then used to model the dynamics of the gas as in step 3 of Section \ref{sec:intro}.

\subsection{Target Galaxy: NGC4697}\label{subsec:data}
To test the above model, we perform dynamical analysis of a well-studied galaxy, NGC4697. This galaxy's black hole has been previously measured [9] using the standard procedure described in Section \ref{sec:intro} via the KinMS dynamical modelling package [14]. To determine the stellar mass component of the galaxy, the authors of that work fit an MGE profile to a Hubble Space Telescope (HST) image. Since NGC4697 contains a low-luminosity AGN [15], the authors in [9] attempt to quantify this potential source of bias in the stellar mass profile by removing the innermost Gaussian component of their fitted MGE. In the remainder of this work, we will refer to the MGE model whose innermost component has been removed as "w/o AGN", and the model whose innermost component was kept as "w/ AGN". 

\subsubsection{Radio interferometry data}
The previous SMBH mass measurement was performed using data from the Atacama Large Millimeter Array (ALMA); we obtain this data from the ALMA archive and use the Common Astronomy Software Applications software for standard data reduction.

The observations made by a radio interferometer are not images, but rather $N(N-1)$ Fourier components of the sky, where $N$ is the number of antennae in the array. Often, these observations (known as visibilities) are transformed into images using the CLEAN algorithm [16], but this algorithm leads to difficulties in estimating uncertainties on the output images. On the other hand, in visibility space, the noise is Gaussian. To estimate the noise variance, we bin the visibilities onto a regular grid using a Kaiser-Bessel window function to minimize aliasing artifacts [17]. Within each grid, we use the Kaiser-Bessel-weighted sample variance to estimate the uncertainty. If the number of visibilities within the grid is less than 5, we expand the grid until this criterion is met. Since we use the sample variance to estimate the uncertainty, the resulting probability distribution for the noise model is Student's T [18].

\subsection{Dynamical modelling}
\label{sec: SuperMAGE}
To jointly infer the stellar mass distribution and the black hole mass in NGC4697, we integrate the neural network from Section \ref{sec:Nuker} into our PyTorch-based dynamical modelling pipeline, SuperMAGE (in prep.). While the details of SuperMAGE are outside the scope of this paper, we provide here an outline of its functionality. SuperMAGE builds a model for emitted light from the gas and combines it with the velocity curve corresponding to the mass profile to generate the predicted spatio-spectral variations in light emitted from gas orbiting the galactic centre. The generated spatio-spectral cube consists of an image of the galaxy for every frequency in the data. Each image is then Fourier transformed and masked to produce a model for the data. SuperMAGE is similar to and tested against KinMS [14] and the mge\_vcirc function of JamPy [19, 20] but unlike those software packages, it is automatically differentiable and can model observations in the Fourier (visibility) plane.

We adopt a Bayesian framework for uncertainty quantification, where the target distribution is 
\begin{equation}\label{eq:Bayes}
    p(\mathbf{x} \mid \mathbf{D}) \propto p(\mathbf{D} \mid \mathbf{f}(\mathbf{x})) p(\mathbf{x}), 
\end{equation}
where $\mathbf{D}$ is the data described in Section \ref{subsec:data}, $\mathbf{f}$ is our model for the galaxy (which consists of SuperMAGE and the Nuker model outlined in Section \ref{sec:Nuker}), and $\mathbf{x}$ are the parameters of our model. While the true likelihood $p(\mathbf{D} \mid \mathbf{f}(\mathbf{x}))$ is Student T (see Section \ref{subsec:data}), we approximate it with a Gaussian distribution with the standard deviation multiplied by a factor of 2. This conservative rescaling is slightly larger than the factor ($\sim 1.77$) needed for a Gaussian's 99\% confidence interval to match that of Student's T for the minimum number of degrees of freedom per grid cell used in Section \ref{subsec:data}. For priors on the model parameters ($p(\mathbf{x})$), we use the broadest possible uniform priors which cover all physically realistic parameter values (for a full list of priors, see Appendix \ref{priors}). We sample the posterior using Metropolis-Adjusted Langevin Dynamics [21]. For details on sampler tuning, see Appendix \ref{sec:appendix_MALA}.

\section{Results and discussion}
\begin{figure}[h]
    \centering
    \includegraphics[width=0.8\linewidth]{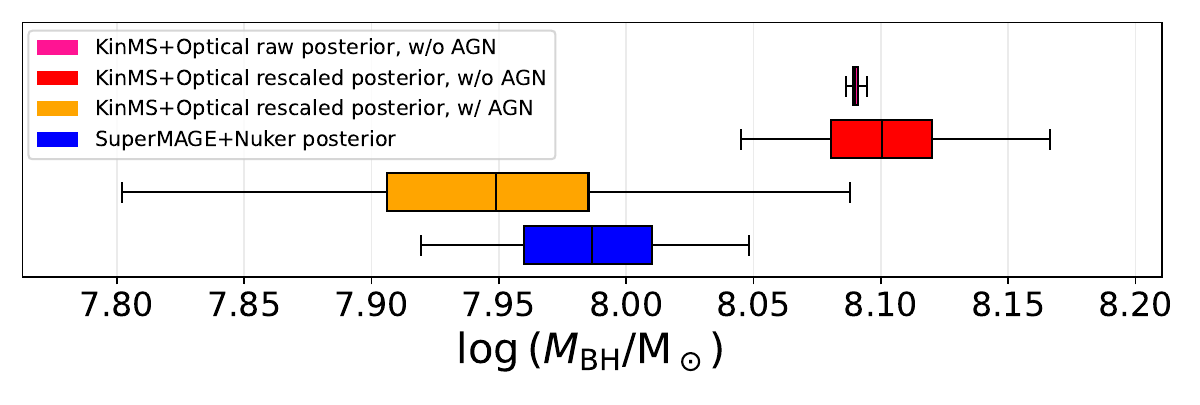}
    \vspace{-0.03\linewidth}
    \caption{Credible intervals for the black hole mass in NGC4697 (box: $1\sigma$, whiskers: $3\sigma$). To produce self-consistent results, KinMS requires an empirically-derived uncertainty rescaling to account for systematic uncertainty (this rescaling has been applied to the second and third box plots). The first two box plots show the KinMS model's inferred SMBH mass given a stellar light profile with the innermost Gaussian removed to account for the presence of light emitted by the AGN ("w/o AGN"); the third box plot shows the KinMS model's inferred black hole mass if the innermost Gaussian is kept as part of the stellar light profile ("w/ AGN").}
    \label{fig:MBH}
    \vspace{-0.01\linewidth}
\end{figure}

Using the Nuker profile as our stellar mass model, our marginalized posterior distribution for the black hole mass yields masses that are broadly consistent with measurements using an MGE model fitted to optical data (Figure \ref{fig:MBH}). We compare our mass measurement to measurements previously obtained using two optical MGE models (see Section \ref{subsec:data} for details): one model whose innermost component has been removed ("w/o AGN") and another whose innermost component was retained ("w/ AGN"). We find that our black hole mass estimate lies between the value from the MGE model "w/o AGN" and the value from the model "w/ AGN". If the true stellar mass distribution of NGC4697 follows our fitted Nuker profile, then our results imply that the AGN correction in [9] was an overcorrection which removed stellar light in addition to AGN light and thereby slightly biased the black hole mass estimate. 

\begin{figure}[h]
    \hspace{-0.2\linewidth}
    \centering
    \includegraphics[width=0.5\linewidth]{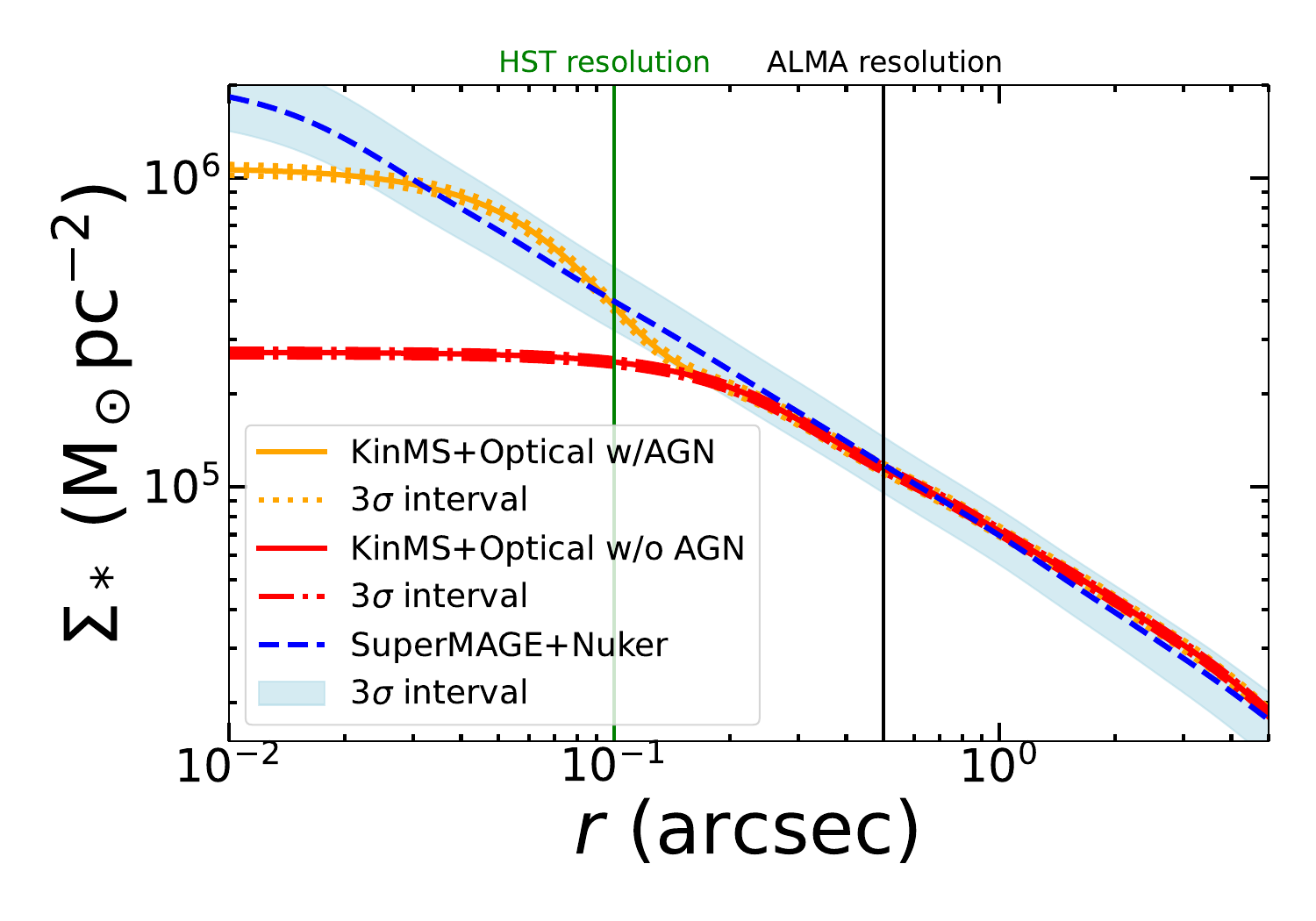}
    \includegraphics[width=0.5\linewidth]{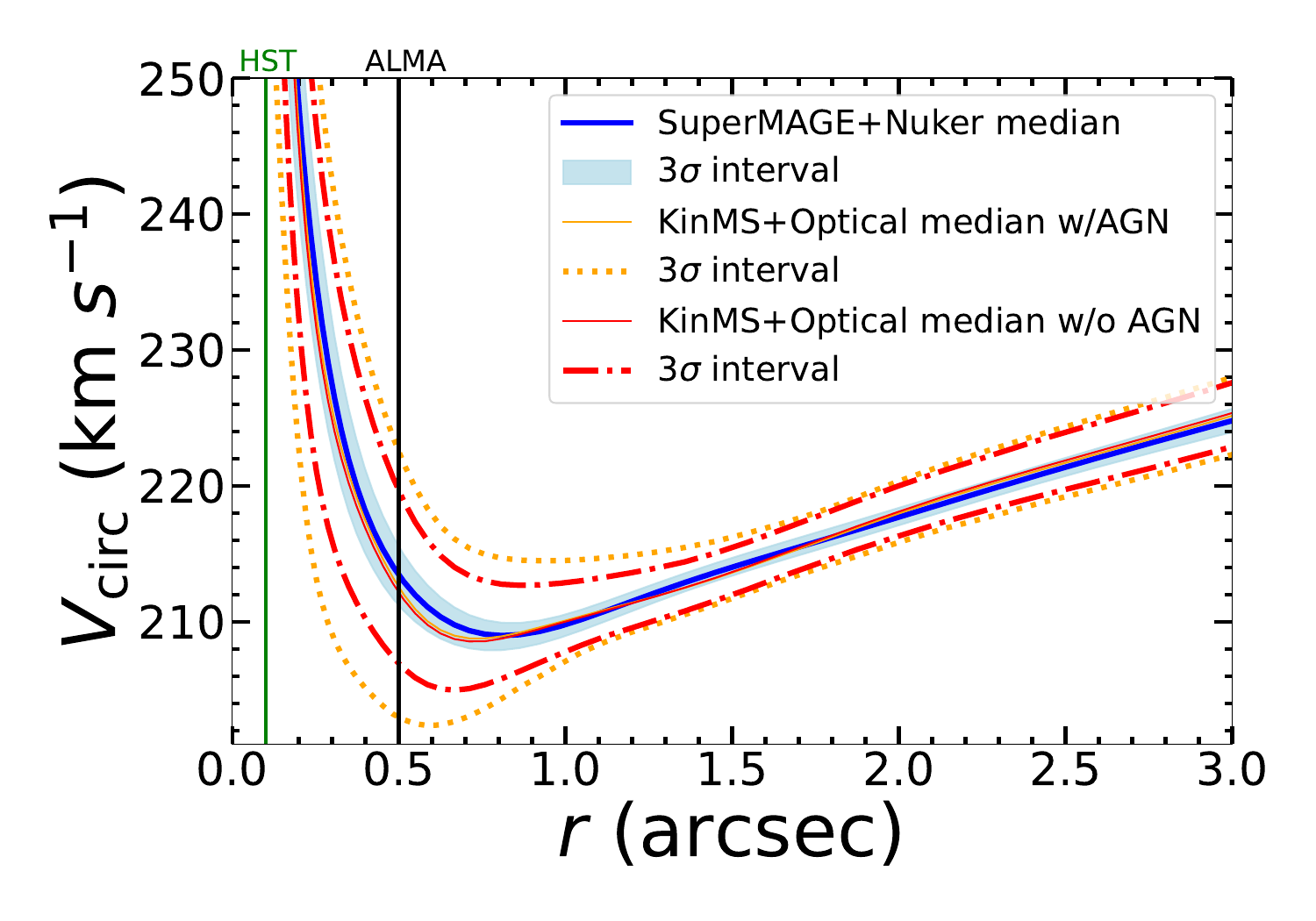}
    \hspace{-0.2\linewidth}
    \vspace{-0.01\linewidth}
    \caption{Left: 2D projected mass density for the optically-fitted MGE profile and our dynamically-fitted Nuker profile. Right: Derived orbital velocity curves for the KinMS+MGE model and our SuperMAGE+Nuker model. The credible intervals for KinMS include the uncertainty rescaling. }
    \label{fig:nukervsmge}
\end{figure}

The effect of stellar mass model choice is further illustrated by the left panel of Figure \ref{fig:nukervsmge}, which compares the best-fit stellar mass density profile using our Nuker model to the MGE profiles. Our inferred Nuker profile follows a power law shape down to 0.02 arcseconds, where it begins to taper due to the width of the innermost Gaussian component in our fixed grid (see Section \ref{sec: neural net}). Meanwhile, the MGE model "w/ AGN" contains a small but statistically insignificant rise, which may be attributed to light from the AGN, before tapering off below the HST resolution limit. The model "w/o AGN" tapers much more rapidly, above the resolution limit of HST but still below the resolution limit of our ALMA data; this rapid taper could be an underestimate of the stellar mass contained within the innermost region of NGC4697. Nevertheless, at radii resolved by the ALMA data, all three models converge to the same values for the stellar mass profile, despite the Nuker model not having been fit to any optical data. 

While the stellar mass models (and hence the inferred black hole masses) diverge at small radii, the total mass profile of the galaxy inferred from ALMA data is consistent across all three methods. This is illustrated in the right panel of Figure \ref{fig:nukervsmge}, which plots the inferred rotational velocity profiles. Our Nuker-based model (using SuperMAGE) has significantly less scatter, which we attribute to an improved fit to the data. The improved fit to the data is due to our model's flexibility: we jointly sample the shape of the stellar mass profile in addition to the other dynamical parameters. This figure demonstrates that the fitted galaxy mass profiles are consistent to within $3\sigma$ between the MGE- and the Nuker-based models, despite the KinMS model having been fit to both optical data and radio data while the Nuker-based model is constrained solely by radio data. 

Compared to previous state-of-the-art dynamical black hole mass modelling, our methodology eliminates the reliance on dust-free optical imaging by instead constraining stellar mass dynamically using only radio data and a neural network-based parametrization of the mass profile. This permits us to fit a wider range of galaxies for which suitable optical imaging is impossible due to the presence of dust or an AGN. Furthermore, our methodology opens the door to performing the first dynamical SMBH mass measurement in a high-redshift gravitationally lensed galaxy, as the limited number of gravitational lenses and the dustiness of high-redshift galaxies make optical data unreliable. 

\section{Acknowledgements}
This work is partially supported by Schmidt Sciences, a philanthropic initiative founded by Eric and Wendy Schmidt as part of the Virtual Institute for Astrophysics (VIA). This work is in part supported by computational resources provided by Calcul Quebec and the Digital Research Alliance of Canada. 
M.J.Y.B. acknowledges support from Mitacs through the Globalink program. M.J.Y.B also acknowledges support from UdeM through the Bourse d’excellence du centenaire, Bourse du passage accéléré, and the Bourse Gilles Beaudet. M.J.Y.B acknowledges technical support from Claude Pascal Duquette and Luc Turbide.
H.Z. acknowledges support from a Science and Technology Facilities Council (STFC) DPhil studentship under grant ST/X508664/1 and the Balliol College J. T. Hamilton Scholarship in physics. M.B. was supported by STFC consolidated grant ‘Astrophysics at Oxford’ ST/K00106X/1 and ST/W000903/1.
Y.H. and L.P. acknowledge support from the Canada Research Chairs Program, the National Sciences and Engineering Council of Canada through grants RGPIN-2020-05073 and 05102.

\newpage
\section*{References}
{
\small

[1] Ferrarese, L., Merritt, D., 2000, ApJ, 539, L9

[2] Gültekin, K. et al., 2009, ApJ, 698, 198

[3] McConnell, N. J., Ma, C.-P., 2013, ApJ, 764, 184

[4] Zhang H., et al., 2024, MNRAS, 530, 3240

[5] Davis, T. A. et al., 2013a, Nature, 494, 328

[6] Ruffa I., et al., 2023, MNRAS, 522, 6170

[7] Cappellari M., 2002, MNRAS, 333, 400

[8] Davidson, J. R. et al., 2024, ApJ, 972, 127

[9] Davis T. A., Bureau M., Onishi K., Cappellari M., Iguchi S., Sarzi M., 2017, MNRAS, 468, 4675

[10]Floyd, D. J. E. et al., 2004, MNRAS, 355, 196

[11] Lauer, T. R. et al., 1995, AJ, 110, 2622

[12]Baes, M., 2020, A\&A, 634, A109







[13] Kingma, D. P. and Ba, J., 2014, arXiv preprint, arXiv:1412.6980

[14] Davis, T. A. et al., 2013b, MNRAS, 429, 534

[15] Wrobel, J.M., Terashima, Y., and Ho, L.C., 2008, ApJ, 675, 1041

[16] Högbom J. A., 1974, A\&AS, 15, 417

[17] Kaiser, J. F., 1966, Systems analysis by digital computer. John Wiley and Sons, New York, NY.

[18] Student, 1908, Biometrika, 6, 1.

[19] Cappellari, M., MNRAS, 2008, 390, 71-86

[20] Cappellari, M., MNRAS, 2020, 494, 4819-4837

[21] Besag, J., 1994, Comments on “Representations of knowledge in complex systems” by Grenander, U. and Miller, M.I., Journal of the Royal Statistical Society Series B, 56, 591

[22] Roberts, G. O., and Rosenthal, J. S., 1998, Journal of the Royal Statistical Society Series B, 60(1), 255
\newpage
\appendix
\section{Training data parameter range}\label{Nuker_range}
We generate the training set on a uniform grid of Nuker shape parameters, $\alpha$, $\gamma$, $\beta-\gamma$, and $r_b$, with 50 values of $\alpha$ ranging from $0.1$ to $10$, 50 values of $\gamma$ ranging from $0.001$ to $1.2$, 50 values of $\beta-\gamma$ ranging from $0.3$ to $3.0$, and 50 values of $r_b$ ranging from 2 parsecs to 2 kiloparsecs. These parameter bounds are determined by typical ranges of Nuker parameters in the literature (e.g., [8]). We apply bounds on $\beta-\gamma$ instead of $\beta$, as physical Nuker profiles require the outer power law slope to be considerably steeper than the inner slope, but do not impose strict constraints on the outer slope itself. For each of the $50^4=6.25e6$ parameter sets, we generate the corresponding Nuker profile $\Sigma(r)$, assuming $\Sigma_b=1$.

\section{Uniform prior ranges}\label{priors}

For our neural network's internal MGE model, we fix the MGE Gaussian widths on a log-uniform grid between $1$ parsec and $10$ kiloparsecs to sample the full range of physical scales of a galaxy, using 64 Gaussians in total. For all other dynamical parameters, we allow values within the following range:
\begin{table}[h]
\centering
\caption{Parameter bounds used in the model.}
\begin{tabular}{lccp{6.8cm}}
\toprule
\textbf{Parameter} & \textbf{Lower} & \textbf{Upper} & \textbf{Description} \\
\midrule
scale\_length & 0.0 & 10.0 & Scale length [arcsecs] \\
$\log_{10} M_{\bullet}/\mathrm{M_\odot}$ & 6.0 & 9.79934 &log SMBH mass; corresponds to $(10^{6}$, $6.3\times10^{9})\,M_{\odot}$ \\
inclination & 72 & 89 & Inclination [deg] \\
qintr & 0.1 & 1.0 & Intrinsic axis ratio \\
alpha & 0.1 & 10.0 & Slope parameter \\
gamma\_minus\_beta & -3.0 & -0.3 & See Appendix \ref{Nuker_range} for discussion \\
gamma & 0.001 & 1.2 & Inner slope \\
break\_r & 0.036 & 5.0 & Break radius [arcsecs] \\
intensity\_rb & 1.0 & 5.0 & Mass density at the break radius [$\mathrm{M_{\odot}}/\mathrm{pc}^2$]\\
PA & 0.0 & 359.9 & Position angle [deg] \\
line\_broadening & 0.1 & 30.0 & Line broadening prior [km/s] \\
systemic\_velocity & 1220.0 & 1260.0 & Systemic velocity [km/s] \\
x & -1.0 & 1.0 & Spatial offset in $x$ [arcsecs] \\
y & -1.0 & 1.0 & Spatial offset in $y$ [arcsecs] \\
Flux & 1.0 & 50.0 & Integrated flux [Jy km/s] \\
\bottomrule
\end{tabular}
\end{table}

\section{Metropolis-Adjusted Langevin Dynamics tuning procedure}\label{sec:appendix_MALA}
We tune the mass matrix by running a series of burn-in stages and recalculating the mass matrix at the best-fit sampled point in each stage using the empirical Fisher Information, which is calculated as described in the following section. We then tune the step size by running a second series of burn-in stages with different step sizes; we keep the step size that yields an acceptance ratio closest to 0.574 [22].
\subsection{Empirical Fisher Matrix}\label{sec:appendix_fisher}
Let the whitened residual be defined as
\[
r(\boldsymbol{\theta}) \;=\; \frac{\hat{\mathbf{y}}(\boldsymbol{\theta}) - \mathbf{y}}{\boldsymbol{\sigma}}
\;\;\;\;\;\;\; \text{with } \mathbf{y} \text{ the data, } 
\hat{\mathbf{y}}(\boldsymbol{\theta}) \text{ the model prediction, and } 
\boldsymbol{\sigma} \text{ the uncertainty.}
\]

Let the Jacobian of the residuals be
\[
J(\boldsymbol{\theta}) \;=\; \frac{\partial r(\boldsymbol{\theta})}{\partial \boldsymbol{\theta}} .
\]

The empirical Fisher information matrix (Gauss–Newton approximation to the Hessian of the log-likelihood) is then
\[
F(\boldsymbol{\theta}) \;=\; J(\boldsymbol{\theta})^{\top} J(\boldsymbol{\theta}) .
\]

For numerical stability, we regularize with a ridge term:
\[
\tilde{F}(\boldsymbol{\theta}) \;=\; F(\boldsymbol{\theta}) \;+\; \lambda I ,
\]
where \(\lambda = \max\!\Bigl(\texttt{ridge\_frac}\cdot \mathrm{median}(\mathrm{diag}(F)), \;\texttt{ridge\_min}\Bigr).\)

Finally, the covariance approximation is given by
\[
\mathrm{Cov}(\boldsymbol{\theta}) \;\approx\; \tilde{F}(\boldsymbol{\theta})^{-1}.
\]

\section{Corner plot for posterior samples}
\begin{figure}[h]
    \centering
    \includegraphics[width=\linewidth]{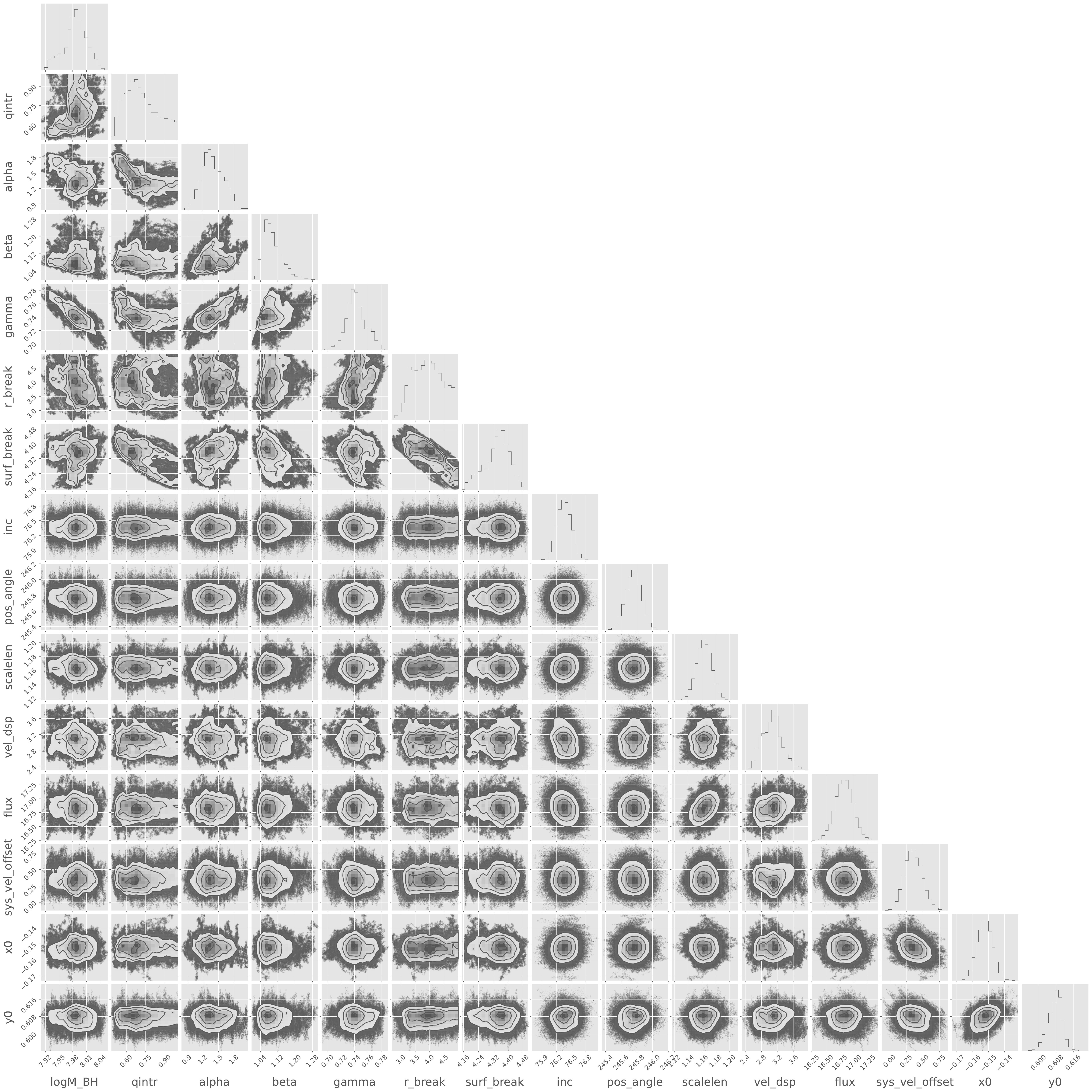}
    \caption{Corner plot for posterior samples for all free parameters in our SuperMAGE+Nuker model. Note that the break radius is constrained to values outside the maximum extent of gas in the galaxy (roughly 3 arcseconds), which indicates that the galaxy is well described by a single power law.}
    \label{fig:placeholder}
\end{figure}

\end{document}